\documentclass[aps,pra,twocolumn,showpacs,superscriptaddress]{revtex4}%
\usepackage{graphics}
\usepackage{amsmath}
\usepackage{graphicx}
\usepackage{amsfonts}
\usepackage{amssymb}
\usepackage{hyperref}
\begin{document}
\title{Modular network for high-rate quantum conferencing}
\author{Carlo Ottaviani}
\email{carlo.ottaviani@york.ac.uk}
\affiliation{Computer Science
and York Centre for Quantum Technologies, University of York, York
YO10 5GH, UK}
\author{Cosmo Lupo}
\affiliation{Department of Physics and Astronomy, University of Sheffield, Hounsfield Road,
Sheffield, S3 7RH, UK}
\author{Riccardo Laurenza}
\affiliation{QSTAR, INO-CNR and LENS, Largo Enrico Fermi 2, 50125 Firenze, Italy}
\author{Stefano Pirandola}
\email{stefano.pirandola@york.ac.uk}
\affiliation{Computer Science and York Centre for Quantum
Technologies, University of York, York YO10 5GH, UK}
\affiliation{Research Laboratory of Electronics, Massachusetts
Institute of Technology, Cambridge, Massachusetts 02139, USA}

\begin{abstract}
One of the main open problems in quantum communication is the
design of efficient quantum-secured networks. This is a
challenging goal, because it requires protocols that guarantee
both unconditional security and high communication rates, while
increasing the number of users. In this scenario,
continuous-variable systems provide an ideal platform where high
rates can be achieved by using off-the-shelf optical components.
At the same time, the measurement-device independent architecture
is also appealing for its feature of removing a substantial
portion of practical weaknesses.  Driven by these ideas, here we
introduce a modular design of continuous-variable network where
each individual module is a measurement-device-independent star
network. In each module, the users send modulated coherent states
to an untrusted relay, creating multipartite secret correlations
via a generalized Bell detection. Using one-time pad between
different modules, the network users may share a quantum-secure
conference key over arbitrary distances at constant rate.
\end{abstract}

\maketitle

Quantum communication~\cite{NiCh,QIbook,Watrous} with continuous
variables (CV) systems~\cite{RMP,BraRMP,Ulrikreview} has attracted
increasing attention over the past years. In particular, quantum
key distribution (QKD) has been a rapidly developing
field~\cite{QKDreview}. Theoretical studies have considered
one-way protocols with coherent
states~\cite{GG02,weedbrook2004noswitching,1way2modes}, thermal
protocols~\cite{filip-th1,weed1,usenkoTH1,weed2,weed2way,usenkoREVIEW},
and two-way protocols \cite{pirs2way,
2way2modes,2way2modes2,QIprot1,flood1}, with a number of
experimental demonstrations~\cite{Grosshans2003b,1D-tobias,
ulrik-Nat-Comm-2012,jouguet2013,Huang-scirep2016,ulrik-entropy,JosephEXP,QIprot2,QIprot3,flood2,flood3}.
It is known that CV-QKD\ protocols may achieve very high rates. As
a matter of fact, ideal coherent-state
protocols~\cite{weedbrook2004noswitching} may achieve rates as
high as half of the Pirandola-Laurenza-Ottaviani-Banchi
bound~\cite{PLOB15,PLOBarxiv} for private communication over a
lossy channel, i.e., $-\log_{2}(1-\eta)$ bits per use, with $\eta$
being the channel transmissivity (see Ref.~\cite{TQCreview} for a
recent review on bounds for private communication).

In addition to point-to-point protocols, there has been effort
towards network
implementations~\cite{Kimble,telereview,HybridINTERNET}. An
important step is the design of a scalable QKD network whose rate
is high enough to compete with the classical infrastructure.
Another feature to achieve is an end-to-end architecture where
middle nodes may be untrusted. The first steps in this direction
were moved in 2012 with the introduction of a swapping protocol
based on an untrusted relay~\cite{MDI1,MDILo}, a technique that
became known as \textquotedblleft measurement-device
independence\textquotedblright\ (MDI), and recently extended to
CV-QKD~\cite{CVMDIQKD,CVMDIQKD-reply,PRA-S,MDI-FS,compo,MDIideal}.
However, until today, MDI protocols have been limited to a small
number of remote users, e.g., $2$ in Refs.~\cite{MDI1}, or $3$ in
Ref.~\cite{Y-Wu}.

In this work we remove these limitations. In particular, we
introduce a modular architecture that combines trusted and
untrusted nodes, as well as quantum and classical communication
methods, allowing secure quantum conferencing among an arbitrary
number of users. At the core of our design there are MDI
star-network modules, interconnected by shared nodes, which can
run one-time pad protocols between different modules. Each module
consists of a central (untrusted) relay performing a general
N-mode Bell detection that allows an arbitrary number of users to
share the same quantum conference key. The security of the
protocol is first proven in the asymptotic limit of many signals
exchanged, and then extended to the composable setting which
incorporates finite-size effects. This modular design is scalable,
because it allows us to increase arbitrarily the number of users
and the achievable distance between them, while maintaining a high
and constant rate. Moreover, it can be implemented using linear
optical elements, and it allows to add extra modules resorting
just on classical communication protocols. From this point of view
the scheme is very flexible and represents a good prototype to be
developed into a large scale CV-QKD network.

\section{Results}

\subsection{\textbf{Modular network for quantum conferencing}}

In our modular architecture, each individual module is a star
network running a multipartite MDI-QKD quantum conferencing
protocol based on the generalization of symmetric
CV-MDI-QKD~\cite{PRA-S}. Each star-network module $M_{i}$ is
labeled by $i=1,\dots,N^{\star}$ and host $N_{i}$ users. The
generic user $k$ in module $M_{i}$ sends bright coherent
states~\cite{RMP} $|\alpha_{k}^{i}\rangle$ to a central untrusted
relay, whose
amplitude $\alpha_{k}^{i}$ is Gaussianly modulated with variance $\mu_{k}^{i}%
$. With no loss of generality we may assume that $\mu_{k}^{i}$ is
the same for any $k$, so that we may associate a single variance
parameter $\mu_{i}$ to module $M_{i}$. The eavesdropping is
assumed to be performed by entangling cloners~\cite{QKDreview}, so
that the link connecting the arbitrary user $k$ to the relay in
module $M_{i}$ is described by two parameters: the transmissivity
$\eta_{k}^{i}$ of the link, and its thermal noise
$\bar{n}_{k}^{i}$.

Within module $M_{i}$\ the untrusted relay performs a multipartite Bell
detection on the incoming $N_{i}$ modes; this consists of a suitable cascade
of beam-splitters followed by $N_{i}$ homodyne detections, as shown in
Fig.~\ref{starPP}. The homodynes on the left measure quadratures $\hat{q}%
_{2},\dots,\hat{q}_{N_{i}}$, while the single one on the right measures
$\hat{p}$. The outcomes of the measurements are combined into the global
outcome $\gamma_{i}:=(q_{2},\ldots,q_{N_{i}},p)$. After $\gamma_{i}$ is
broadcast to the users of the module, their individual variables $\alpha
_{k}^{i}$ share correlations that can be post-processed into a secret key
$K_{i}$ via classical error correction and privacy amplification. All the
users in module $M_{i}$ reconcile their data with respect to a trusted user
which is shared with another module $M_{j}$.

\begin{figure}[ptb]
\vspace{-0.0cm}
\par
\begin{center}
\includegraphics[width=0.50\textwidth]{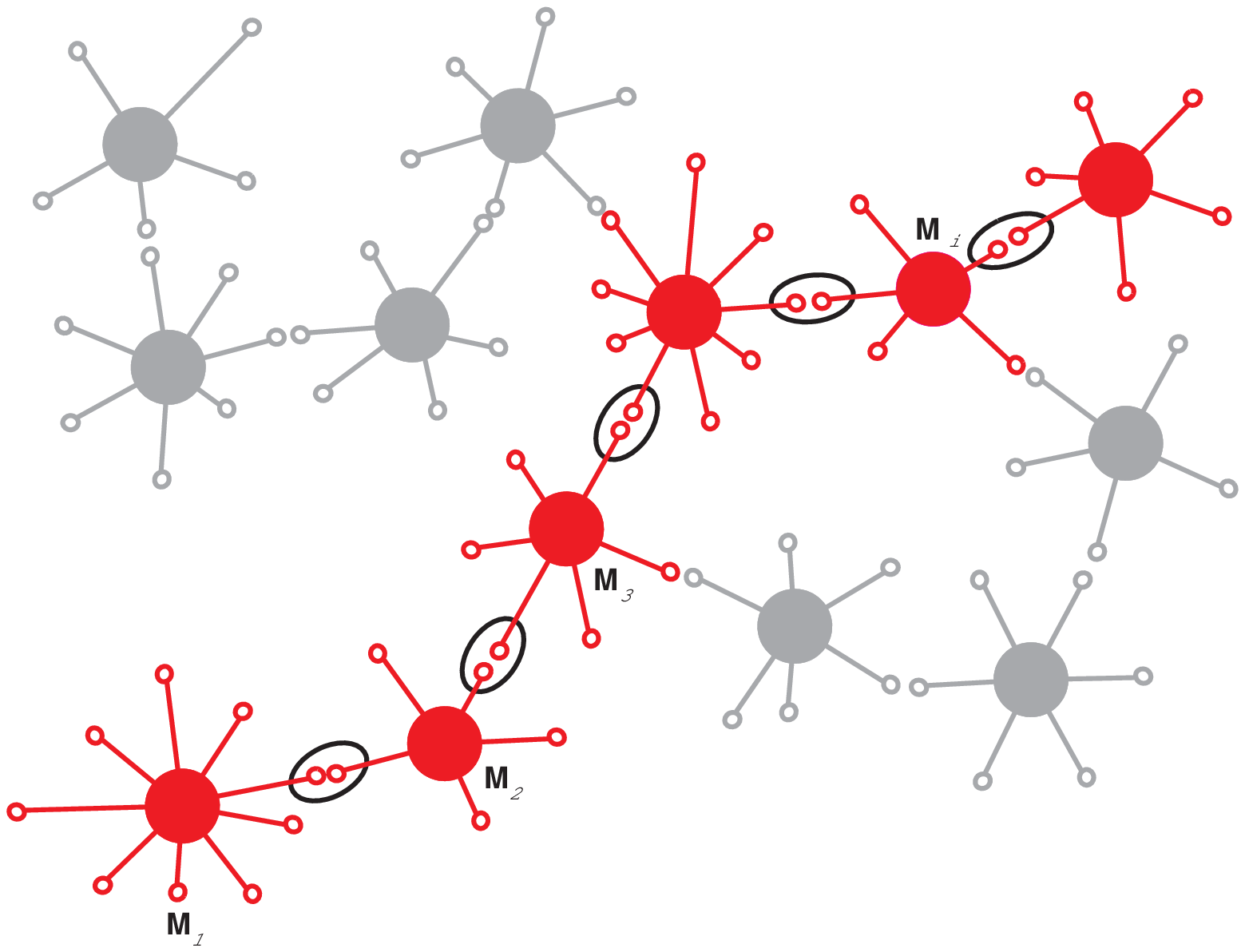} \vspace{-0.8cm}
\end{center}
\par
\vspace{-0.1cm}\caption{Modular network for secure quantum conferencing. Each
module $M_{i}$ is a continuous-variable (CV) measurement-device-independent
(MDI) quantum key--distribution (QKD) star network, composed by a central
untrusted relay and $N_{i}$ trusted users, whose connections are independently
affected by loss and noise. Each module $M_{i}$ first runs an independent
protocol of quantum conference key-agreement. Because two different modules
have a shared trusted user, we may implement classical one-time pad sessions
where the keys from each module are processed into a final common key for the
entire network.}%
\label{Modular}%
\end{figure}

To reduce the parameters of the problem we may introduce the minimum
transmissivity $\eta_{i}=\min_{k\in\lbrack1,N_{i}]}\eta_{k}^{i}$ and maximum
thermal noise $\bar{n}_{i}=\max_{k\in\lbrack1,N_{i}]}\bar{n}_{k}^{i}$
associated to module $M_{i}$. From a physical point of view this is a
symmetrization of the star network to the worst-case scenario, assuming all
its links to have the worst combination of parameters. This condition clearly
provides a lower bound $K(\mu_{i},N_{i},\eta_{i},\bar{n}_{i})$ to the actual
key rate $K_{i}$ of the module. By optimizing over the Gaussian modulation, we
may consider the value $K(N_{i},\eta_{i},\bar{n}_{i}):=\max_{\mu_{i}%
}K(\mu_{i},N_{i},\eta_{i},\bar{n}_{i})$. Once each module has
generated its key, the shared nodes run sessions of one-time pad
where the keys from different modules are composed to generate a
common key for all the network, with rate
\begin{equation}
K_{\text{net}}=\min_{i\in\lbrack1,N^{\star}]}K(N_{i},\eta_{i},\bar{n}_{i}).
\label{GLOBAL-KEY}%
\end{equation}
Therefore, the entire network can work at the rate of the least
performing module. However, if this rate is high then the
lego-like structure of the network allows all the users to
communicate at the same high rate no matter how far they are from
each other (see Fig.~\ref{Modular}).

\subsection{\textbf{Detailed description the MDI star-network module}}

\begin{figure}[ptb]
\vspace{-0.0cm}
\par
\begin{center}
\includegraphics[width=0.48\textwidth]{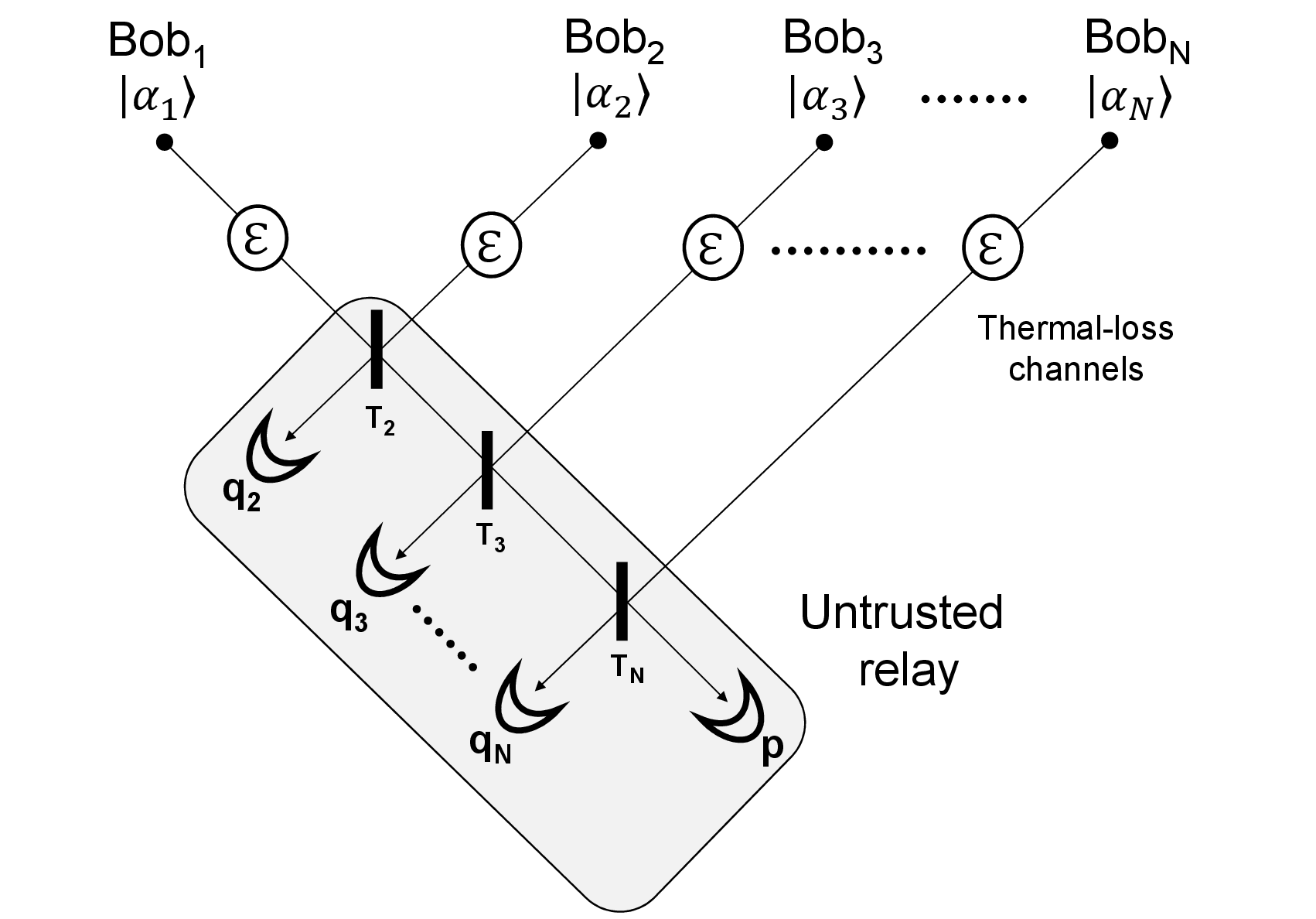}
\end{center}
\par
\vspace{-0.4cm}\caption{Each Bob sends a Gaussian-modulated
coherent state $\left\vert \alpha_{k}\right\rangle $ to an
untrusted relay through a link which is described by a
thermal-loss channel $\mathcal{E}$ with transmissivity $\eta$ and
thermal noise $\bar{n}$. At the relay, the incoming states are
subject to a multipartite continuous-variable (CV) Bell detection,
by using a cascade of beam-splitters with transmissivities
$T_{k}=1-k^{-1}$ for $k=2,\ldots,N$, followed by homodyne
detectors in the $\hat{q}$\ or $\hat{p}$
quadrature as shown in the figure. The global outcome $\gamma=(q_{2}%
,\ldots,q_{N},p)$ is broadcast to the Bobs, so that a posteriori correlations
are created in their local variables $\alpha_{1},\ldots,\alpha_{N}$. These
correlations are used to extract a secret key for quantum conferencing.}%
\label{starPP}%
\end{figure}

In this section we describe the modus operandi of a single module.
To simplify the notation we omit the label $i$. In a single
module, we consider an arbitrary number $N$ of users (or
\textquotedblleft Bobs\textquotedblright) sending
Gaussian-modulated coherent states $|\alpha_{k}\rangle$ to a
middle untrusted relay, as depicted in Fig.~\ref{starPP}. Each of
the coherent states is affected by a thermal-loss channel
$\mathcal{E}$ modeled as a beam-splitter with transmissivity
$\eta$ and thermal noise $\bar{n}$, i.e., mixing the incoming
signal with an environmental thermal state with $\bar{n}$ mean
photons. As explained before, we assume the worst-case scenario,
so that $\eta$ is the minimum transmissivity of the links and
$\bar{n}$ is the maximum thermal noise. After the action of the
channel $\mathcal{E}$\ on each link, the states are detected by a
multipartite $N$-mode Bell detection.

This detection consists of a suitable cascade of beam-splitters followed by
$N$ homodyne detections. More precisely, we have a sequence of beam-splitters
with increasing transmissivities $T_{k}=1-k^{-1}$ for $k=2,\ldots,N$ as
depicted in Fig.~\ref{starPP}. Then, all the homodynes at the left measure the
$\hat{q}$-quadrature while the final one at the bottom measures the $\hat{p}%
$-quadrature, with global outcome $\gamma:=(q_{2},\ldots,q_{N},p)$
(see Supplementary Notes 1 and 2 for further description). One can
check that this measurement ideally projects onto a displaced
version of an asymptotic bosonic state $\Psi$ that realizes the
multipartite Einstein-Podolsky-Rosen
(EPR) conditions $%
{\textstyle\sum\nolimits_{k=1}^{N}}
\hat{p}_{k}=0$ and $\hat{q}_{k}-\hat{q}_{k^{\prime}}=0$ for any $k,k^{\prime
}=1,\ldots,N$.

After the classical outcome $\gamma$ is broadcast to the users,
their individual variables $\alpha_{k}$ will share correlations
which can be post-processed into secret keys via error correction
and privacy amplification. We may pick the shared trusted user as
the one encoding the key, with all the others decoding it in
direct reconciliation~\cite{QKDreview}.

\subsection{\textbf{Entanglement-based representation}}

Let us write the network in entanglement-based representation. For
each user, the coherent state $|\alpha\rangle$ can be generated by
using a two-mode squeezed vacuum (TMSV) state $\Phi_{AB}$ where
mode $B$ is subject to heterodyne detection. The random outcome
$\beta$ of the detection is fully equivalent to prepare a coherent
state on mode $A$ whose amplitude $\alpha $\ is one-to-one with
$\beta$~~\cite{CVMDIQKD}. Recall that a TMSV state is a Gaussian
state with covariance matrix (CM)~\cite{RMP}
\begin{equation}
\mathbf{V}_{AB}=\left(
\begin{array}
[c]{cc}%
\mu\mathbf{I} & \sqrt{\mu^{2}-1}\mathbf{Z}\\
\sqrt{\mu^{2}-1}\mathbf{Z} & \mu\mathbf{I}%
\end{array}
\right)  ,~\left\{
\begin{array}
[c]{l}%
\mathbf{Z}:=\mathrm{diag}(1,-1),\\
\mathbf{I}:=\mathrm{diag}(1,1),
\end{array}
\right.
\end{equation}
where parameter $\mu\geq1$ quantifies the noise variance of each
thermal mode. Up to factors~\cite{CVMDIQKD},\ parameter $\mu$ also
provides the variance of the Gaussian modulation of the coherent
amplitude $\alpha$ on mode $A$ after heterodyning $B$.

Assume that the users have $N$ copies of the same TMSV state, whose $A$-part
is sent to the relay through a communication channel $\mathcal{E}$. Also
assume that the CM of the two-mode state after the channel has the form
\begin{equation}
\mathbf{V}_{AB}^{\prime}=\left(
\begin{array}
[c]{cc}%
x\mathbf{I} & z\mathbf{Z}\\
z\mathbf{Z} & y\mathbf{I}%
\end{array}
\right)  \,.
\end{equation}
Because we consider a thermal-loss channel with transmissivity $\eta$ and
thermal noise $\bar{n}$, we have $x=\eta\mu+(1-\eta)(2\bar{n}+1)$, $y=\mu$,
and $c=\sqrt{\eta}\sqrt{\mu^{2}-1}$. Then, after the Bell measurement and the
communication of the outcome $\gamma$, the local modes $\mathbf{B}:=B_{1}%
$\ldots$B_{N}$ are projected onto a symmetric $N$-mode Gaussian state with CM
\begin{equation}
\mathbf{V}_{\mathbf{B}|\gamma}=\left(
\begin{array}
[c]{cccc}%
\boldsymbol{\Delta} & \boldsymbol{\Gamma} & \cdots & \boldsymbol{\Gamma}\\
\boldsymbol{\Gamma} & \boldsymbol{\Delta} & \ddots & \boldsymbol{\Gamma}\\
\vdots & \ddots & \ddots & \vdots\\
\boldsymbol{\Gamma} & \boldsymbol{\Gamma} & \cdots & \boldsymbol{\Delta}%
\end{array}
\right)  ~, \label{SYMM-CM}%
\end{equation}
where we have set $\boldsymbol{\Gamma}:=(N^{-1}x^{-1}z^{2})\mathbf{Z}$, and%
\begin{equation}
\boldsymbol{\Delta}:=\mathrm{diag}\left(  y-\frac{N-1}{N}\frac{z^{2}}%
{x},y-\frac{1}{N}\frac{z^{2}}{x}\right)  ~.
\end{equation}
Details on the the derivation of Eq.~(\ref{SYMM-CM}) are given in
the Supplementary Note 2.

Note that the conditional state $\rho_{B_{i}B_{j}|\gamma}$ between any pair of
Bobs $i$ and $j$ is Gaussian with CM%
\begin{equation}
\mathbf{V}_{B_{i}B_{j}|\gamma}=\left(
\begin{array}
[c]{cc}%
\boldsymbol{\Delta} & \boldsymbol{\Gamma}\\
\boldsymbol{\Gamma} & \boldsymbol{\Delta}%
\end{array}
\right)  . \label{reduced1}%
\end{equation}
For $N=2$ this state describes the shared state in a standard CV-MDI-QKD
protocol~\cite{CVMDIQKD}. Assuming no thermal noise ($\bar{n}=0$), the state
$\rho_{B_{i}B_{j}|\gamma}$ is always entangled and we may compute its relative
entropy of entanglement (REE) $E_{R}(\rho_{B_{1}B_{2}|\gamma})$%
~\cite{RMPrelent,VedFORMm,Pleniom} using the formula for the relative entropy
between Gaussian states~\cite{PLOB15,fidelityBAN}. This REE provides an upper
bound to the rate achievable by any MDI-QKD protocol (DV or CV) based on a
passive untrusted relay. For $N>2$, one can check that the bipartite state
$\rho_{B_{i}B_{j}|\gamma}$ may become separable when we decrease the
transmissivity $\eta$, while it certainly remains
discordant~\cite{ParisD,GerryD,PirD}. In the multi-user scenario, the security
between two Bobs may still hold because the purification of their state is
held partially by Eve and partially by the other Bobs, which play the role of
trusted noise. In trusted noise QKD we know that security does not rely on the
presence of bipartite entanglement while quantum discord provides a necessary
condition~\cite{discordQKD}.

\begin{figure*}[ptb]
\vspace{-0.0cm}
\par
\begin{center}
\includegraphics[width=0.9\textwidth]{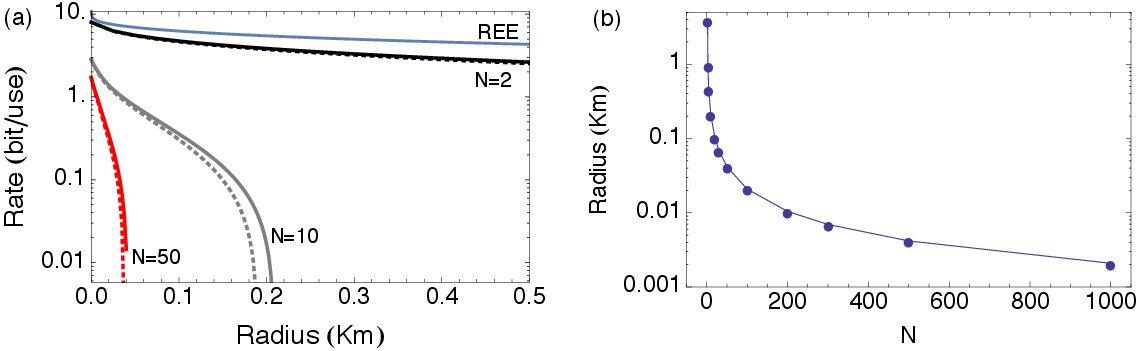}
\end{center}
\par
\vspace{-0.4cm}\caption{Secret key rates and maximum distances for quantum
conferencing. (a)~We plot the conferencing key rate for a
measurement-device-independent star network of $N=2$ (black), $10$ (grey) and
$50$\textbf{\ }(red) users, as a function of the fiber distance $d$ and
assuming thermal noise $\bar{n}=0$ (solid curves) and $\bar{n}=0.05$ (dashed
curves). The top blue curve is the relative entropy of entanglement (REE) of
the reduced bipartite state specified by Eq.~(\ref{reduced1}), which upper
bounds the maximal key rate achievable with standard continuous-variable
measurement-device-independent quantum-key-distribution CV-MDI-QKD ($N=2$).
(b)~We plot the maximum fiber distance $d$ versus the number of users $N$ in a
quantum conference.}%
\label{star}%
\end{figure*}

\subsection{\textbf{Key rate of a star-network module}}

Once $\gamma$ is received, the $i$th Bob heterodynes his local mode $B_{i}$
with random outcome $\beta_{i}$, which is one-to-one with an encoded amplitude
$\alpha_{i}$ in the prepare and measure description. In this way the local
mode $B_{j}$ of the $j$th Bob is mapped into a Gaussian state $\rho
_{\mathbf{B}_{j}|\gamma\beta_{i}}$ with CM $\mathbf{V}_{\mathbf{B}_{j}%
|\gamma\beta_{i}}$
that can be computed using tools from
Refs.~\cite{RMP,detec,detec2,Belldetection}. The subsequent heterodyne
detection of mode $B_{j}$ generates an outcome $\beta_{j}$ which is one-to-one
with an encoded $\alpha_{j}$. It is clear that the Bell detection at the relay
and the local heterodyne meaurements of the various Bobs all commute, so that
we may change their time order in the security analysis of the protocol. Thus,
we can derive the mutual information $I(\beta_{i}:\beta_{j})$ between the two
Bobs. Similarly, we may compute the Holevo information $\chi(\beta
_{i}:\mathbf{E})$ between the $i$th Bob and an eavesdropper (Eve) performing a
collective Gaussian attack~\cite{coll1,coll2,coll3} associated with the
thermal-loss channels~\cite{RMP}.

The expression $K=I(\beta_{i}:\beta_{j})-\chi(\beta_{i}:\mathbf{E})$, which is
a function of all the parameters of the protocol, provides the asymptotic rate
of secret key generation between any pair of users (see Supplementary Note 3).
This is a conferencing key shared among all Bobs; it can be optimized over
$\mu$, and will depend on the number of Bobs $N$ besides the channel
parameters $\eta$ and $\bar{n}$. Assuming a standard optical fiber with
attenuation of $0.2$dB per km, we can map the transmissivity into a fiber
distance $d$, using $\eta:=10^{-0.02d}$. Therefore, we have a rate of the form
$K(\mu,N,d,\bar{n})$ which can be optimized over $\mu$ to give the maximal
conferencing rate $K(N,d,\bar{n})$. The maximization over $\mu$ is required
because for $N>2$ the maximum rate is not obtained in the limit $\mu\gg1$.

The conferencing rate is plotted in Fig.~\ref{star}(a) for an MDI star network
with an increasing number of users $N$. We compare the rates over the
link-distance $d$ for different values of thermal noise $\bar{n}$. As expected
the rate decreases for increasing $N$. Despite this effect, our result shows
that high-rate quantum conferencing is possible. For instance, in a star
network with $N=50$ users at about $d\simeq40$m from the central relay and
thermal noise $\bar{n}=0.05$, the key rate is greater than $\simeq0.1$ bits
per use.
In Fig.~\ref{star}(b), we set $\bar{n}=0$ and plot the maximum distance for
quantum conferencing versus the number of users, solving the equation
$K(N,d,0)=0$. We see a trade-off between maximum distance and number of users.
Despite this trade-off, we conclude that fiber-optic secure quantum
conferencing between tens of users (belonging to a single star-network module)
is indeed feasible within the typical perimeter of a large building.

\subsection{\textbf{Finite-size composable security }}

Within a module, consider a pair of Bobs, $i$ and $j$, with local
variables $\beta_{i}$ and $\beta_{j}$ after heterodyne detection.
They aim at generating a secret key by reconciliating on
$\beta_{i}$. The error correction routine is characterized by an
error correction efficiency
$\xi\in(0,1)$~\cite{Lin015,miliceviv017}, a residual probability
of error $\delta_{\mathrm{EC}}$, and an abort probability $1-p>0$.
We also remark that $\beta_{i}$ must be mapped into a discrete
variable $\bar{\beta}_{i}$ taking $2^{d}$ values per quadrature.

Consider the mutual information $I(\beta_{i}:\beta_{j})$ and Eve's
Holevo bound $\chi(\beta_{i}:\mathbf{E})$ obtained from the
reduced state of Eq.~(\ref{reduced1}). Then, we have the following
estimate for the $\delta$-secret key rate after $n$ uses of the
module~\cite{compo}
\begin{equation}
r_{n}^{\delta}\gtrsim\xi I(\beta_{i}:\beta_{j})-\chi(\beta_{i}:\mathbf{E}%
)-\frac{1}{\sqrt{n}}\,\Delta_{\mathrm{AEP}}(2p\delta_{\mathrm{s}}/3,d)\,,
\label{fsr}%
\end{equation}
where
$\delta=\delta_{\mathrm{s}}+\delta_{\mathrm{EC}}+\delta_{\mathrm{PE}}$,
and
$\Delta_{\mathrm{AEP}}(\xi,d)\leq4(d+1)\sqrt{\log{(2/\xi^{2})}}$.
Here the error term $\delta_{\mathrm{s}}$ is the smoothing
parameter of the smooth conditional min-entropy~\cite{compo}. For
the Holevo bound $\chi(\beta _{i}:\mathbf{E})$ we assume the
worst-case value compatible with the experimental data, up to a
probability smaller than $\delta_{\mathrm{PE}}$. The rate
$r_{n}^{\delta}$ is obtained conditioned that the protocol does
not abort and yields a $\delta$-secure key against collective
Gaussian attacks.

\begin{figure}[ptb]
\vspace{-0.0cm}
\par
\begin{center}
\includegraphics[width=0.45\textwidth]{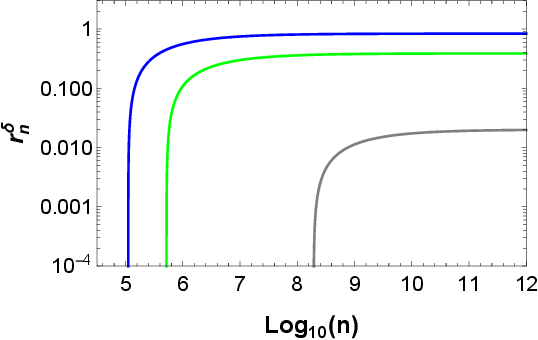}
\end{center}
\par
\vspace{-0.0cm}\caption{Finite-size composable secret key rate
$r_{n}^{\delta}$ of Eq.~(\ref{fsr}) (bits per use) versus sample
size $n$, for error correction efficiency
$\xi=98\%$~\cite{Lin015,miliceviv017}, success probability of the
error correction routine $p=0.9$, and security parameter
$\delta<10^{-20}$. The plot is obtained for a symmetric
configuration with $N$ users at a distance of $0.18$km from the
relay (assuming $0.2$dB loss per km and thermal noise of
$\bar{n}=0.05$ photons). From top to bottom we consider $N=3,5,10$.}%
\label{fig:comp}%
\end{figure}

The generalization to coherent attacks is obtained, as in
Ref.~\cite{compo}, by applying a Gaussian de Finetti reduction. We
find that the asymptotic rates are approximately achieved for
block sizes of $10^{6}-10^{9}$ data points depending on the loss
and noise in the channels. Examples for $N=3,5,10$ are described
in Fig.~\ref{fig:comp}.

The composable security analysis starts from a parameter
estimation procedure. As mentioned above, in estimating the
channel parameters, we adopt the worst-case scenario where we
choose the largest possible value of the thermal noise in each
channel, and the lower available transmissivity, within the
confidence intervals. This procedure may not be optimal at high
loss, so that our estimates for the achievable communication
distances are conservative estimates. Alternative (more
performing) parameter estimation procedures, as those described in
Refs.~\cite{Ruppert-FS,Thearle-FS}, could be adapted to our
network model and further improve its performance.

\section{Discussion}

We have introduced a network for quantum conferencing where
modules can be linked together to achieve constant high-rate
secure communication over arbitrarily long distances. The design
of each module is based on a CV-MDI star network with many users.
Our analysis shows how the secret key-rate of each star network
decreases by increasing the distance from the central relay and/or
the number of users. In ideal conditions, we find that $50$ users
may privately communicate at more than $0.1$ bit per use within a
radius of $40$m, distance typical of a large building. With a
clock of $25$MHz~\cite{Clock}, this is a key rate of the order of
$2.5$Mbits per second for all the users. The secret keys
established in the modules are then cascaded through the entire
modular network: Adjacent modules are connected by trusted users
generating a common key via one-time pad sessions.

We have studied the implementation of our protocol using coherent
states, which are important for practical reasons. In
Supplementary Note~4, we have also considered the case where the
users use squeezed states. In such a case the performance
improves, both in terms of achievable distance and the number of
users. In any case, we remark that our network protocol provides a
powerful application of CV-MDI-QKD that greatly outperforms its DV
counterpart. In fact, while our protocol is deterministic, any
linear optical implementation of a DV multi-partite Bell detection
is highly probabilistic, with a probability of success scaling as
$\simeq2^{-N}$ for $N$ users. This means that a corresponding
DV-MDI-QKD star network has an exponentially low rate no matter at
what distance is implemented. See Supplementary Note 5 for
details.

In Supplementary Note 6, we further analyze the performance of the
protocol based on coherent states, assuming practical
imperfections affecting the homodyne detectors and the beam
splitters of the relay. Such undesirable features may arise from
imperfect beam splitting operations, perturbing the multipartite
Bell detection and degrading the performance of the scheme. Our
results show that a proof-of-principle experiment is feasible with
current technology, allowing to secure up to $9$ users per module,
over a radius of $12$m.

In conclusion, let us remark that our network is based on a generalization of
CV MDI-QKD, where the multiuser keys are extracted within each single module.
For this reason the final key is shorter than the private key extractable by
just two remote end-users. In other words, the key rate of our protocol cannot
reach the existing upper bounds for end-to-end network quantum
communication~\cite{netpublished,networkPIRS} (see also
Refs.~\cite{Azuma2016,Rigo2018}). Finally, let us also note that additional
studies may consider the multimode nature of sources and detectors,
particularly for the optimized configuration of the protocol described in the
Supplementary Note 4. In such a case, the security of the scheme can be
recovered by applying the symmetrization of the multimode source described in
Ref.~\cite{usenko-multimode}.

\section*{Acknowledgments}

This work was supported by the EPSRC via the `UK Quantum
Communications Hub' (EP/M013472/1), the European Union's Horizon
2020 research and innovation program under grant agreement No
820466 (CiViQ), and the Innovation Fund Denmark via the Quantum
Innovation Center Qubiz.

\newpage

\widetext

%
%
%

\begin{center}
\Huge Supplementary Notes
\end{center}

\setcounter{section}{0} \setcounter{equation}{0}
\setcounter{figure}{0} \setcounter{page}{1}
\makeatletter\renewcommand{\theequation}{S\arabic{equation}}
\renewcommand{\thefigure}{S\arabic{figure}}

\section{Supplementary Note 1: Network in the entanglement-based
representation}

We study the security of our multipartite protocol adopting the
entanglement-based (EB) representation, described in Supplementary
Figure.~\ref{starEB}.\begin{figure*}[pth] \vspace{-0.0cm}
\par
\begin{center}
\includegraphics[width=0.5\textwidth]{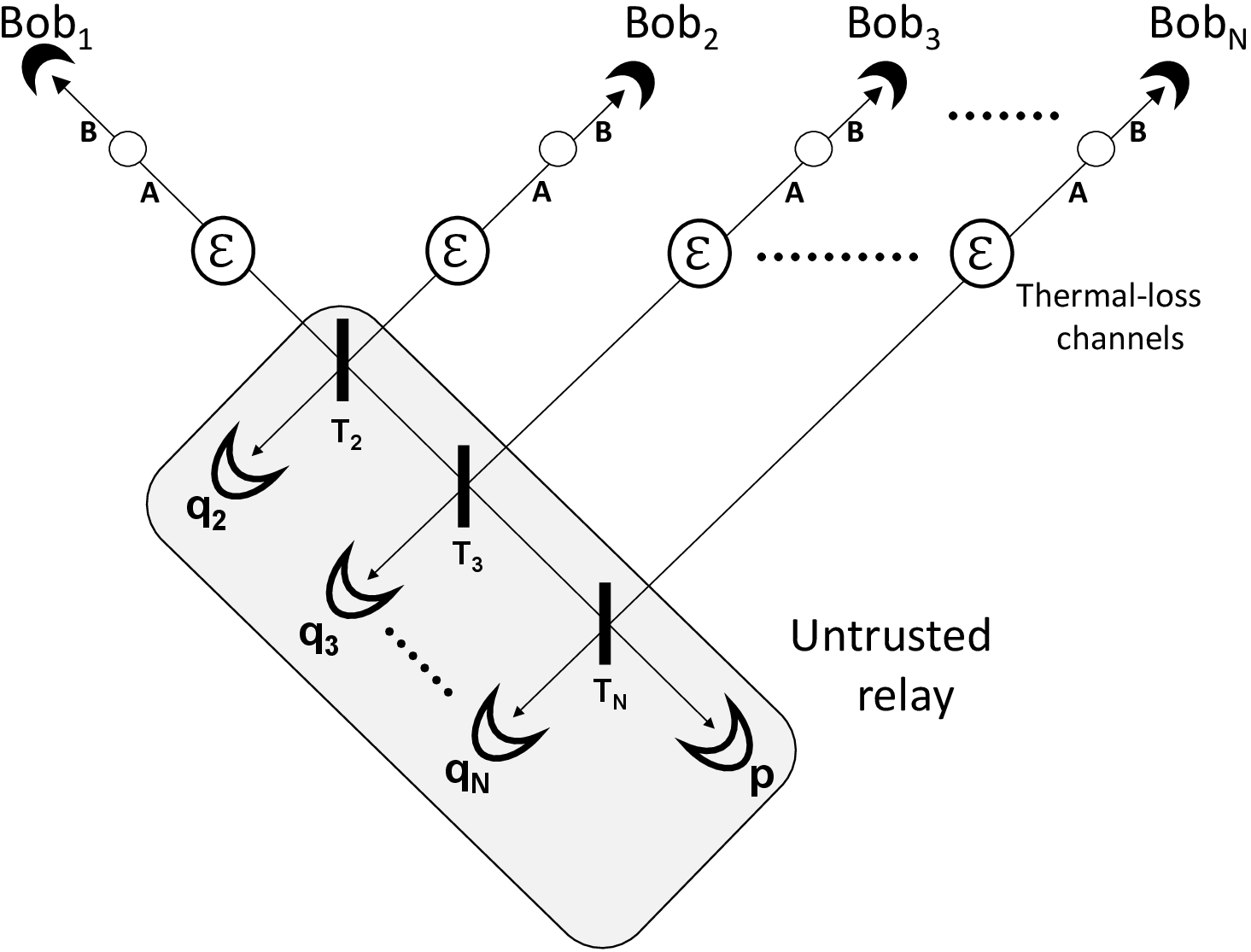}
\end{center}
\par
\vspace{-0.1cm}\caption{Each Bob prepares a two-mode squeezed
vacuum (TMSV) state $\Phi_{AB}$, composed by local mode $B$ and
the traveling mode $A$. The former is kept and measured, while the
latter is sent through the communication channel to the untrusted
relay, where the multipartite Bell detection is performed. In the
squeezed protocol, conditional local squeezings
are applied before Bobs' heterodynes.}%
\label{starEB}%
\end{figure*}
 Each Bob prepares a two-mode squeezed vacuum
state (TMSV) $\Phi_{AB}$ where mode $B$ is kept while mode $A$ is
sent to the relay. Using the EB representation we may then
consider two setups of the network.

\begin{itemize}
\item In a coherent-state configuration of the protocol, each Bob applies
heterodyne detection on his local mode $B$, effectively projecting
the traveling mode $A$ into a Gaussian-modulated coherent state.
This is the practical protocol presented in our main text, which
can be easily done in prepare and measure.

\item In a squeezed configuration of the protocol, the heterodyne detection on
$B$ is performed after the application of a local squeezing, so
that a displaced squeezed state is remotely generated on mode $A$.
Because the optimal local squeezing depends on the multipartite
state shared by the Bobs after the Bell detection, this version of
the protocol needs to be performed in the EB representation.
\end{itemize}

\noindent In the following Notes, we start by describing the
mathematics of the multipartite Bell detection. Then, we consider
the coherent-state protocol and the squeezed-state protocol. Both
of them are studied from the point of view of quantum
conferencing.

\section{Supplementary Note 2: Multipartite Bell detection\label{BellSEC}}

In the main text we introduce a multipartite Bell measurement
where $N$ input modes first pass through an interferometer
composed of $N$ cascaded beam-splitters, and then are measured by
homodyne detections with global outcome $\gamma$. In EB
representation, the effect of the multipartite Bell detection is
to distribute a conditional $N$-mode Gaussian state to the network
users. Here we compute the covariance matrix (CM) of such $N$-mode
conditional state. We start by showing a convenient dual
representation for the multimode Bell detection, where the
interferometer on $A$ modes can be replaced by a conjugate one
acting on the $B$ modes. Then we compute the action of the
homodyne detector on the $A$ modes, and the symplectic
transformation of the $B$ modes.

\subsection{Dual representation of the multipartite Bell detection\label{Ax}}

Consider a system of $N$ pairs of bosonic modes, where the
quadrature vector of the $k^{\text{th}}$ pair is denoted as
\begin{equation}
\xi_{k}=(\hat{q}_{k}^{A},\hat{p}_{k}^{A},\hat{q}_{k}^{B},\hat{p}_{k}^{B})^{T},
\end{equation}
for $k=1,\dots,N$ (this pair is in the hands of the
$k^{\text{th}}$ Bob). In a symmetric setting all the pairs of
modes are prepared in the same state, which we assume to be a
zero-mean Gaussian state with CM
\begin{equation}
\mathbf{V}_{k}=\left(
\begin{array}
[c]{cccc}%
x & 0 & z & 0\\
0 & x & 0 & -z\\
z & 0 & y & 0\\
0 & -z & 0 & y
\end{array}
\right)  \,.\label{CM2}%
\end{equation}
To represent the system of $2N$ modes, we define the quadrature
vector
\begin{align}
\xi &  =(\xi^{A},\xi^{B})^{T}\\
&  =(\hat{q}_{1}^{A},\hat{q}_{2}^{A},\dots,\hat{q}_{N}^{A},\hat{p}_{1}^{A}%
,\hat{p}_{2}^{A},\dots,\hat{p}_{N}^{A},\hat{q}_{1}^{B},\hat{q}_{2}^{B}%
,\dots,\hat{q}_{N}^{B},\hat{p}_{1}^{B},\hat{p}_{2}^{B},\dots,\hat{p}_{N}%
^{B})^{T}\,,
\end{align}
so that the CM of the multimode state reads
\begin{equation}
\mathbf{V}=\left(
\begin{array}
[c]{cccc}%
x\mathbf{I}_{N} & 0 & z\mathbf{I}_{N} & 0\\
0 & x\mathbf{I}_{N} & 0 & -z\mathbf{I}_{N}\\
z\mathbf{I}_{N} & 0 & y\mathbf{I}_{N} & 0\\
0 & -z\mathbf{I}_{N} & 0 & y\mathbf{I}_{N}%
\end{array}
\right)  \,,\label{spec}%
\end{equation}
where $\mathbf{I}_{N}$ is the $N\times N$ identity matrix. Now we
describe the action of the interferometer on this input state. The
interferometer is described by a symplectic transformation on the
$A$ modes. Such a transformation maps the vector of $A$ quadrature
\begin{equation}
\xi^{A}=(\hat{q}_{1}^{A},\hat{q}_{2}^{A},\dots,\hat{q}_{N}^{A},\hat{p}_{1}%
^{A},\hat{p}_{2}^{A},\dots,\hat{p}_{N}^{A})^{T}%
\end{equation}
as follows
\begin{equation}
\xi^{A}\rightarrow\left(
\begin{array}
[c]{cc}%
\mathbf{R} & 0\\
0 & \mathbf{R}%
\end{array}
\right)  \xi^{A}\,,
\end{equation}
where $\mathbf{R}$ is a $N\times N$ orthogonal matrix. This
diagonal form follows from the fact that the beam-splitter
transformations are chosen in such a way that they do not mix
$\hat{q}$'s and $\hat{p}$'s. The
transformation for the CM is%
\begin{equation}
\mathbf{V}\rightarrow\mathbf{V}^{\prime}=\left(
\begin{array}
[c]{cccc}%
\mathbf{R} & 0 & 0 & 0\\
0 & \mathbf{R} & 0 & 0\\
0 & 0 & \mathbf{I}_{N} & 0\\
0 & 0 & 0 & \mathbf{I}_{N}%
\end{array}
\right)  \left(
\begin{array}
[c]{cccc}%
x\mathbf{I}_{N} & 0 & z\mathbf{I}_{N} & 0\\
0 & x\mathbf{I}_{N} & 0 & -z\mathbf{I}_{N}\\
z\mathbf{I}_{N} & 0 & y\mathbf{I}_{N} & 0\\
0 & -z\mathbf{I}_{N} & 0 & y\mathbf{I}_{N}%
\end{array}
\right)  \ \left(
\begin{array}
[c]{cccc}%
\mathbf{R}^{T} & 0 & 0 & 0\\
0 & \mathbf{R}^{T} & 0 & 0\\
0 & 0 & \mathbf{I}_{N} & 0\\
0 & 0 & 0 & \mathbf{I}_{N}%
\end{array}
\right)  \,.
\end{equation}
Applying $\mathbf{R}^{T}\mathbf{R}=\mathbf{RR}^{T}=\mathbf{I}_{N}$
to the specific form of the CM in Eq.~(\ref{spec}) we may write
\begin{align}
\mathbf{V}^{\prime}  &  =\left(
\begin{array}
[c]{cccc}%
\mathbf{I}_{N} & 0 & 0 & 0\\
0 & \mathbf{I}_{N} & 0 & 0\\
0 & 0 & \mathbf{R}^{T} & 0\\
0 & 0 & 0 & \mathbf{R}^{T}%
\end{array}
\right)  \left(
\begin{array}
[c]{cccc}%
\mathbf{R} & 0 & 0 & 0\\
0 & \mathbf{R} & 0 & 0\\
0 & 0 & \mathbf{R} & 0\\
0 & 0 & 0 & \mathbf{R}%
\end{array}
\right)  \left(
\begin{array}
[c]{cccc}%
x\mathbf{I}_{N} & 0 & z\mathbf{I}_{N} & 0\\
0 & x\mathbf{I}_{N} & 0 & -z\mathbf{I}_{N}\\
z\mathbf{I}_{N} & 0 & y\mathbf{I}_{N} & 0\\
0 & -z\mathbf{I}_{N} & 0 & y\mathbf{I}_{N}%
\end{array}
\right)  \left(
\begin{array}
[c]{cccc}%
\mathbf{R}^{T} & 0 & 0 & 0\\
0 & \mathbf{R}^{T} & 0 & 0\\
0 & 0 & \mathbf{R}^{T} & 0\\
0 & 0 & 0 & \mathbf{R}^{T}%
\end{array}
\right)  \left(
\begin{array}
[c]{cccc}%
\mathbf{I}_{N} & 0 & 0 & 0\\
0 & \mathbf{I}_{N} & 0 & 0\\
0 & 0 & \mathbf{R} & 0\\
0 & 0 & 0 & \mathbf{R}
\end{array}
\right) \nonumber\\
&  =\left(
\begin{array}
[c]{cccc}%
\mathbf{I}_{N} & 0 & 0 & 0\\
0 & \mathbf{I}_{N} & 0 & 0\\
0 & 0 & \mathbf{R}^{T} & 0\\
0 & 0 & 0 & \mathbf{R}^{T}%
\end{array}
\right)  \left(
\begin{array}
[c]{cccc}%
x\mathbf{I}_{N} & 0 & z\mathbf{I}_{N} & 0\\
0 & x\mathbf{I}_{N} & 0 & -z\mathbf{I}_{N}\\
z\mathbf{I}_{N} & 0 & y\mathbf{I}_{N} & 0\\
0 & -z\mathbf{I}_{N} & 0 & y\mathbf{I}_{N}%
\end{array}
\right)  \left(
\begin{array}
[c]{cccc}%
\mathbf{I}_{N} & 0 & 0 & 0\\
0 & \mathbf{I}_{N} & 0 & 0\\
0 & 0 & \mathbf{R} & 0\\
0 & 0 & 0 & \mathbf{R}%
\end{array}
\right)  \,.
\end{align}
The meaning of this last equation is that the state obtained by
passing the $A$ modes through the interferometer described by the
matrix $\mathbf{R}$ is the same that would be obtained by passing
the $B$ modes through a conjugate interferometer described by the
matrix $\mathbf{R}^{T}$. We have therefore found an equivalent
dual representation for the multipartite Bell detection: Instead
of first passing the $A$ mode through the interferometer and then
measure them, we can equivalently first measure the $A$ modes and
then pass the $B$ modes through the conjugate interferometer. Of
course this dual representation is valid as long as the total
input CM takes the specific form in Eq.~(\ref{spec}).

\subsection{Homodyne detections\label{Bx}}

Let us assume the dual representation of the multipartite Bell
detection. First we need to consider homodyne detection on the $A$
modes. In our scheme only one mode (mode $A_{1}$) is measured in
the quadrature $\hat{p}$, while all the others are measured in the
quadrature $\hat{q}$. Consider a pair of modes described by the CM
in Eq.~(\ref{CM2}). If the $A$ mode is homodyned in the $\hat{q}$
quadrature, the resulting conditional CM of the $B$ mode is
\begin{equation}
\left(
\begin{array}
[c]{cc}%
y-z^{2}/x & 0\\
0 & y
\end{array}
\right)  \,.
\end{equation}
If instead the $\hat{p}$ quadrature is measured we obtain the
conditional CM
\begin{equation}
\left(
\begin{array}
[c]{cc}%
y & 0\\
0 & y-z^{2}/x
\end{array}
\right)  \,.
\end{equation}
Therefore, in terms of the vector of quadratures $\xi^{B}=(\hat{q}_{1}%
^{B},\hat{q}_{2}^{B},\dots,\hat{q}_{N}^{B},\hat{p}_{1}^{B},\hat{p}_{2}%
^{B},\dots,\hat{p}_{N}^{B})^{T}$ the CM of the $B$ modes,
conditioned on the measurement output $\gamma$ on the $A$ modes,
is
\begin{align}
\mathbf{V}_{\mathbf{B}|\gamma}^{\text{in}}  &  =\left(
\begin{array}
[c]{cc}%
\mathbf{V}_{Q} & 0\\
0 & \mathbf{V}_{P}%
\end{array}
\right)  ,\\
\mathbf{V}_{Q}  &  :=\left(
\begin{array}
[c]{cc}%
y & 0\\
0 & (y-z^{2}/x)\mathbf{I}_{N-1}%
\end{array}
\right)  ,\\
\mathbf{V}_{P}  &  :=\left(
\begin{array}
[c]{cc}%
y-z^{2}/x & 0\\
0 & y\mathbf{I}_{N-1}%
\end{array}
\right)  .
\end{align}
This is the conditional CM of the $B$\ modes of the Bobs before
the action of the conjugate interferometer (note that the Bobs
know the first moments of their local reduced states from the
value of the Bell outcome $\gamma$).

\subsection{Action of the (conjugate) interferometer\label{Cx}}

The network of cascaded beam-splitters with transmissivities
$T_{k}=1-1/k$, for $k=1,\dots,N$, as introduced in the main body
of the paper, is described by the following linear transformations
on the position quadratures
\begin{align}
\hat{q}_{1}  &  \rightarrow\frac{1}{\sqrt{N}}\sum_{j=1}^{N}\hat{q}_{j},\\
\hat{q}_{k}  &  \rightarrow\sqrt{1-k^{-1}}\left(  \hat{q}_{k}-\frac{1}%
{k-1}\sum\limits_{i=1}^{k-1}\hat{q}_{i}\right)
~~\text{for~}k=2,\dots,N,
\end{align}
together with analogous transformations on the momentum
quadratures $\hat {p}_{j}$'s. We then obtain the elements of the
symplectic matrix $\mathbf{R}$
\begin{align}
R_{1j}  &  =\frac{1}{\sqrt{N}}~,\\
R_{kj}  &  =-\frac{1}{\sqrt{k(k-1)}}~\text{~for~}k=2,\dots,N-1\text{,}\\
R_{kk}  &  =\sqrt{1-k^{-1}}.
\end{align}
From the interferometer $\mathbf{R}$ we derive the conjugate interferometer $\mathbf{R}^{T}%
$\ and we compute the CM of the $B$ modes from
$\mathbf{V}_{\mathbf{B}|\gamma }^{\text{in}}$, finding
\begin{equation}
\mathbf{V}_{\mathbf{B}|\gamma}:=\left(
\begin{array}
[c]{cc}%
\mathbf{V}_{Q}^{\prime} & 0\\
0 & \mathbf{V}_{P}^{\prime}%
\end{array}
\right)  =\left(
\begin{array}
[c]{cc}%
\mathbf{R}^{T}\mathbf{V}_{Q}\mathbf{R} & 0\\
0 & \mathbf{R}^{T}\mathbf{V}_{P}\mathbf{R}%
\end{array}
\right)  .
\end{equation}
After simple algebra we find the following elements (for $i,j=1,\dots,N$)%
\begin{align}
\left(  \mathbf{V}_{Q}^{\prime}\right)  _{ij}  &  =\sum_{k,l}R_{ki}{V_{Q}%
}_{kl}R_{lj}\\
&  =yR_{1i}R_{1j}+\left(  y-\frac{z^{2}}{x}\right)  \sum_{k=2}^{N}R_{ki}%
R_{kj}\\
&  =\frac{z^{2}}{x}\,R_{1i}R_{1j}+\left(  y-\frac{z^{2}}{x}\right)
\sum
_{k=1}^{N}R_{ki}R_{kj}\\
&  =\frac{z^{2}}{Nx}+\left(  y-\frac{z^{2}}{x}\right)
\delta_{ij}\,,
\end{align}
and similarly
\begin{equation}
\left(  \mathbf{V}_{P}^{\prime}\right)  _{ij}=\sum_{k,l}R_{ki}{V_{P}}%
_{kl}R_{lj}=-\frac{z^{2}}{Nx}+y\,\delta_{ij}\,.
\end{equation}
Thus, we have computed the conditional CM of Bobs' $B$ modes after
the multipartite Bell measurement with outcome $\gamma$. The
resulting state $\rho_{\mathbf{B}|\gamma}$ is symmetric under
permutation of the $B$ modes, and the correlation between any pair
of modes scales as $1/N$. In particular, we may re-write this CM
as follows
\begin{equation}
\mathbf{V}_{\mathbf{B}|\gamma}=\left(
\begin{array}
[c]{cccc}%
\boldsymbol{\Delta} & \boldsymbol{\Gamma} & \dots & \boldsymbol{\Gamma}\\
\boldsymbol{\Gamma} & \boldsymbol{\Delta} & \ddots & \boldsymbol{\Gamma}\\
\vdots & \ddots & \ddots & \vdots\\
\boldsymbol{\Gamma} & \boldsymbol{\Gamma} & \dots & \boldsymbol{\Delta}%
\end{array}
\right)  \,,\label{V2sym}%
\end{equation}
with%
\begin{equation}
\boldsymbol{\Gamma}=\left(
\begin{array}
[c]{cc}%
\frac{z^{2}}{Nx} & 0\\
0 & -\frac{z^{2}}{Nx}%
\end{array}
\right)  \,,\label{V2B}%
\end{equation}
and
\begin{equation}
\boldsymbol{\Delta}=\left(
\begin{array}
[c]{cc}%
y-\frac{N-1}{N}\frac{z^{2}}{x} & 0\\
0 & y-\frac{z^{2}}{Nx}%
\end{array}
\right)  .\label{V2A}%
\end{equation}
Now consider the $i^{\text{th}}$ and $j^{\text{th}}$ Bobs with
quadrature
vector $\xi_{ij}=(\hat{q}_{i}^{B},\hat{p}_{i}^{B},\hat{q}_{j}^{B},\hat{p}%
_{j}^{B})^{T}$. It is easy to check that the conditional (reduced)
CM reads
\begin{align}
\mathbf{V}_{B_{i}B_{j}|\gamma}  &  =\left(
\begin{array}
[c]{cc}%
\boldsymbol{\Delta} & \boldsymbol{\Gamma}\\
\boldsymbol{\Gamma} & \boldsymbol{\Delta}%
\end{array}
\right) \nonumber\\
&  =\left(
\begin{array}
[c]{cccc}%
y-\frac{N-1}{N}\frac{z^{2}}{x} & 0 & \frac{z^{2}}{Nx} & 0\\
0 & y-\frac{z^{2}}{Nx} & 0 & -\frac{z^{2}}{Nx}\\
\frac{z^{2}}{Nx} & 0 & y-\frac{N-1}{N}\frac{z^{2}}{x} & 0\\
0 & -\frac{z^{2}}{Nx} & 0 & y-\frac{z^{2}}{Nx}%
\end{array}
\right)  \,.\label{V2}%
\end{align}

\section{Supplementary Note 3: Coherent-state protocol\label{coherentSEC}}

\subsection{Holevo bound}

In order to compute Eve's Holevo information we exploit the fact
that $\rho_{\mathbf{EB}|\gamma}$ is pure, where $\mathbf{E}$ are
Eve's output modes. We also use the fact that, after the
heterodyne of the $i^{\text{th}}$
Bob with outcome $\beta_{i}$, the conditional state $\rho_{\mathbf{EB}%
_{i}|\gamma\beta_{i}}$ with $\mathbf{B}_{i}:=B_{1},..,B_{i-1},B_{i+1}%
,..,B_{N}$ is pure. Therefore, we may write the Holevo bound as
\begin{equation}
\chi(\beta_{i}:\mathbf{E})=S(\rho_{\mathbf{E}|\gamma})-S(\rho_{\mathbf{E}%
|\gamma\beta_{i}})=S(\rho_{\mathbf{B}|\gamma})-S(\rho_{\mathbf{B}_{i}%
|\gamma\beta_{i}})~,
\end{equation}
where $S$ is the von Neumann entropy. These entropic terms can
therefore be computed from Bobs' CM $V_{\mathbf{B}|\gamma}$ of
Eq.$~$(\ref{V2sym}). This has a single $N$-degenerate symplectic
eigenvalues, given by the following expression
\begin{equation}
\nu=\sqrt{y\left(  y-\frac{z^{2}}{x}\right)  }\,,\label{SI-NI-T}%
\end{equation}
where we defined
\begin{equation}
x:=\eta\mu+(1-\eta)\omega,~~y:=\mu\,,~~z:=\sqrt{\eta(\mu^{2}-1)}%
.\label{xyzDEF}%
\end{equation}
From Eq.~(\ref{SI-NI-T}) we obtain the total von Neumann entropy,
given by
\begin{equation}
S(\rho_{\mathbf{B}|\gamma})=Nh(\nu),\label{SI-von-neuman-tot}%
\end{equation}
where
\begin{equation}
h(x):=\frac{x+1}{2}\log\frac{x+1}{2}-\frac{x-1}{2}\log\frac{x-1}{2}~.
\end{equation}
Then we compute the symplectic spectrum of the conditional CM $V_{\mathbf{B}%
_{i}|\gamma\beta_{i}}$. This corresponds to state $\rho_{\mathbf{B}_{i}%
|\gamma\beta_{i}}$, which describes the state of the $N-1$ users
conditioned to heterodyne detection on one arbitrary Bob. The
doubly conditional CM is
$\mathbf{V}_{\mathbf{B}_{i}|\gamma\beta_{i}}$, and has $N-2 $ \
identical symplectic eigenvalues given by Eq.~(\ref{SI-NI-T}), and
one given by the following expression
\begin{equation}
\nu_{N}=\sqrt{\frac{\lambda\bar{\lambda}}{\tau\bar{\tau}}},\label{nik}%
\end{equation}
where
\begin{align}
\lambda &  :=N\omega\mu+\eta\left[  1+\left(  N-1-N\omega\right)
\mu\right]
,\\
\bar{\lambda}  &  :=N\omega\mu+\eta\left[  N-1-\left(
N\omega-1\right)
\mu\right]  ,\\
\tau &  :=N\omega(1-\eta)+\eta(N-1+\mu),\\
\bar{\tau}  &  :=N\omega(1-\eta)+\eta\left[  (N-1)\mu+1\right]  ~.
\end{align}
Then we may compute the following conditional von Neumann entropy
\begin{equation}
S(\rho_{\mathbf{B}_{i}|\gamma\beta_{i}})=(N-2)h(\nu)+h(\nu_{N})~,\label{Scon}%
\end{equation}
which, together with Eq.~(\ref{SI-von-neuman-tot}), gives the
following Holevo bound
\begin{equation}
\chi(\beta_{i}:\mathbf{E})=2h(\nu)-h(\nu_{N}).\label{chi}%
\end{equation}

\subsection{Mutual information}

The state of two given users is described by the CM of
Eq.~(\ref{V2}). To compute the mutual information between two Bob,
we need the conditional CM after local heterodyne measurement,
i.e., we may apply the formula for heterodyne detection to CM
$\mathbf{V}_{B_{i}B_{j}|\gamma}$, obtaining
\begin{equation}
\mathbf{V}_{B_{i}|\gamma\beta_{j}}=\boldsymbol{\Delta}-\boldsymbol{\Gamma
}\left[  \boldsymbol{\Delta}+\mathbf{I}\right]
^{-1}\boldsymbol{\Gamma}~,
\end{equation}
from which it is simple to compute the mutual information between
the two users, which is given by
\begin{equation}
I(\beta_{i}:\beta_{j})=\frac{1}{2}\log_{2}\frac{\sigma_{\text{s}}}%
{\sigma_{\text{n}}},\label{MI}%
\end{equation}
with
\begin{align}
\sigma_{\text{s}}  &  =1+\det\mathbf{V}_{B_{i}|\gamma}+\text{Tr}%
\mathbf{V}_{B_{i}|\gamma},\nonumber\\
\sigma_{\text{n}}  &  =1+\det\mathbf{V}_{B_{j}|\gamma\beta_{i}}+\text{Tr}%
\mathbf{V}_{B_{j}|\gamma\beta_{i}}~.\label{sigmas}%
\end{align}
It is interesting to study the behavior of
$I(\beta_{i}:\beta_{j})$ in terms of the number of users $N$ and
the Gaussian modulation $\mu$. We can check that, for $N=2$ and
large modulation ($\mu\gg1$), one has
\begin{equation}
\sigma_{\text{n}}^{N=2}=\frac{4\left[  \omega(1-\eta)+\eta\right]  ^{2}}%
{\eta^{2}},
\end{equation}
which recovers the result of Ref.~\cite{PRA-S} for standard
CV-MDI-QKD. More generally, for $N>2$, we find
\begin{equation}
\sigma_{\text{n}}^{N>2}=\frac{2\left(  N-2\right)  \left[
\omega(1-\eta
)+\eta\right]  }{\left(  N-1\right)  \eta}\mu+f(\eta,\omega,N),\label{sigman}%
\end{equation}
where $f(\eta,\omega,N)$ is a function that does not depend on the
modulation $\mu$. Because $\sigma_{\text{n}}^{N>2}$ has a linear
dependence in $\mu$, the key rate and the achievable distance
decrease for increasing $\mu$. This means that an optimal $\mu$
needs to be identified in terms of the other parameters. We
implicitly assume this optimization in the computation of the key
rate
\begin{equation}
K=I(\beta_{i}:\beta_{j})-\chi(\beta_{i}:\mathbf{E})~.
\end{equation}

\section{Supplementary Note 4: Squeezed protocol\label{SqueezedSEC}}

\subsection{Local squeezing}

Better key rates in terms of bits per use are obtained if the
users apply local active operations, namely squeezing, before
their heterodyne detections.
Let us consider the reduced CM $\mathbf{V}_{B_{i}B_{j}|\gamma}=\mathbf{V}%
_{B_{i}B_{j}|\gamma}(x,y,z,N)$ in Eq.~(\ref{V2}), and define the
following parameters
\begin{equation}
s:=\sqrt{\frac{y-\kappa}{y-\kappa\left(  N-1\right)  }},~~\kappa:=\frac{z^{2}%
}{Nx}.\label{kappadef}%
\end{equation}
The local squeezing operations are chosen in such a way to
transform
$\mathbf{V}_{B_{i}B_{j}|\gamma}$ into%
\begin{equation}
\mathbf{W}_{B_{i}B_{j}|\gamma}=\left(
\begin{array}
[c]{cc}%
\boldsymbol{\alpha} & \boldsymbol{\epsilon}\\
\boldsymbol{\epsilon} & \boldsymbol{\alpha}%
\end{array}
\right)  ,\label{Vij}%
\end{equation}
where
\begin{equation}
\boldsymbol{\alpha}=\sqrt{(y-\kappa)[y-\kappa\left(  N-1\right)  ]}%
\mathbf{I},~~\boldsymbol{\epsilon}=\kappa~\mathrm{diag}(s,s^{-1})\,\mathbf{Z},
\end{equation}
with $\mathbf{I}=\mathrm{diag}(1,1)$ and
$\mathbf{Z}=\mathrm{diag}(1,-1)$. Assuming that all Bobs perform
such local operations, the total conditional CM
$\mathbf{V}_{\mathbf{B}|\gamma}$ becomes
\begin{equation}
\mathbf{W}_{\mathbf{B}|\gamma}=\left(
\begin{array}
[c]{cccc}%
\boldsymbol{\alpha} & \boldsymbol{\epsilon} & \cdots & \boldsymbol{\epsilon}\\
\boldsymbol{\epsilon} & \boldsymbol{\alpha} & \ddots & \boldsymbol{\epsilon}\\
\vdots & \ddots & \ddots & \vdots\\
\boldsymbol{\epsilon} & \boldsymbol{\epsilon} & \cdots & \boldsymbol{\alpha}%
\end{array}
\right)  .\label{SI-Vtot}%
\end{equation}
Now assume the propagation of the $A$ modes through thermal-loss
channels with equal transmissivity $\eta$ and equal thermal
variance $\omega=2\bar{n}+1$ with $\bar{n}$ the mean number of
photons. Then, we may write $x=\eta
\mu+(1-\eta)\omega$, $y=\mu$, $z=\sqrt{\eta(\mu^{2}-1)}$ and therefore%
\begin{equation}
\kappa=\frac{\eta(\mu^{2}-1)}{N[\eta\mu+(1-\eta)\omega]}.\label{KAPPA}%
\end{equation}
It is clear that the local squeezings are conditional, i.e., they
are fully determined once the CM~$\mathbf{V}_{B_{i}B_{j}|\gamma}$\
is known to the Bobs. This means that they need to retain their
local modes $B$ until after parameter estimation. Once the CM is
known they may then apply their local operations. Also note that
these operations cannot be simulated in the classical
post-processing because they are not passive but active
(squeezing). For this reason this version of the protocol can only
be performed in EB representation and not in prepare and measure.

\subsection{Holevo bound}

For the computation of the Holevo function, the only change occurs
on the conditional von Neumann entropy. We then compute the
symplectic spectrum of the CM
$\mathbf{W}_{\mathbf{B}_{i}|\gamma\beta_{i}}$ of the state $\tilde
{\rho}_{\mathbf{B}_{i}|\gamma\beta_{i}}$. We find the previous
symplectic eigenvalue $\nu$\ of Eq.~(\ref{SI-NI-T}) with $N-2$
multiplicity, and the eigenvalue
\begin{equation}
\tilde{\nu}_{N}=\frac{\mu(\mu-N\kappa)+\sqrt{(\mu-\kappa)(\mu-\left(
N-1\right)  \kappa)}}{1+\sqrt{(\mu-\kappa)(\mu-\left(  N-1\right)  \kappa)}%
}.\label{SI-nik}%
\end{equation}
Therefore, we may write
\begin{equation}
S(\tilde{\rho}_{\mathbf{B}_{i}|\gamma\beta_{i}})=(N-2)h(\nu)+h(\tilde{\nu}%
_{N})~,\label{SI-Scon}%
\end{equation}
and we derive
\begin{equation}
\chi(\beta_{i}:\mathbf{E})=2h(\nu)-h(\tilde{\nu}_{N}).\label{SI-chi}%
\end{equation}
As expected this result does not depend on $i$.

\subsection{Mutual information}

The mutual information is computed using the CM $\mathbf{W}_{B_{i}B_{j}%
|\gamma}$ in Eq.~(\ref{Vij}).
\begin{figure*}[th]
\vspace{-0.0cm}
\par
\begin{center}
\includegraphics[width=0.95\textwidth]{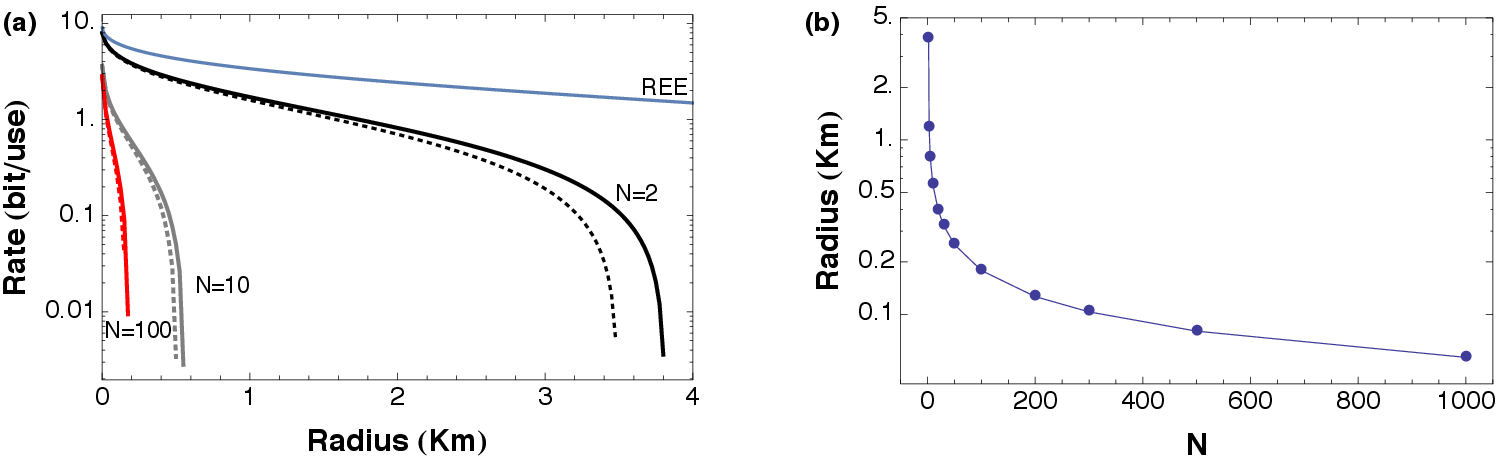}
\end{center}
\par
\vspace{-0.0cm}\caption{Performance of the squeezed protocol,
which is performed in the entanglement-based (EB) representation,
where the various Bobs perform conditional local squeezings on
their $B$-modes before detection. We plot the key rates and the
maximum distances for quantum conferencing in panels (a) and (b).
In~(a)~we plot the conferencing key rate for $N=2$ (black), $10$
(grey) and $100$ (red) users, as a function of the fiber distance
$d$ from the relay. We assumed an attenuation of $0.2$dB/Km, and
thermal noise $\bar{n}=0$ (solid curves) and $\bar{n}=0.05$
(dashed curves). The top blue curve is the upper bound provided by
the relative entropy of entanglement (REE) of the reduced
bipartite state specified by Eq.~(\ref{V2}). In panel~(b)~we plot
the maximum fiber distance $d$ from the relay versus the number of
users $N$. From panel (a), we see that the use of squeezed states
allows to improve the performance. The scheme allows to achieve a
key rate of $0.1$ per use for $100$ users over a radius of $200$m.}%
\label{figS}%
\end{figure*}
Heterodyning mode $B_{i}$, we derive%
\begin{equation}
\mathbf{W}_{B_{j}|\gamma\beta_{i}}=\boldsymbol{\alpha}-\boldsymbol{\epsilon
}\left[  \boldsymbol{\alpha}+\mathbf{I}\right]
^{-1}\boldsymbol{\epsilon}~.
\end{equation}
Therefore, we may compute the mutual information as~\cite{CVMDIQKD}%
\begin{equation}
I(\beta_{i}:\beta_{j})=\frac{1}{2}\log_{2}\frac{\sigma_{\text{s}}}%
{\sigma_{\text{n}}},\label{SI-I}%
\end{equation}
where%
\begin{align}
\sigma_{\text{s}} &  =1+\det\mathbf{W}_{B_{i}|\gamma}+\text{Tr}\mathbf{W}%
_{B_{i}|\gamma},\nonumber\\
\sigma_{\text{n}} &  =1+\det\mathbf{W}_{B_{j}|\gamma\beta_{i}}+\text{Tr}%
\mathbf{W}_{B_{j}|\gamma\beta_{i}}~.\label{SI-sigma}%
\end{align}
As previously mentioned, the optimal $\mu$ needs to be identified
in terms of the other parameters, and then used to compute the key
rate
\begin{equation}
K=I(\beta_{i}:\beta_{j})-\chi(\beta_{i}:\mathbf{E})~.\label{K-squeezed}%
\end{equation}
The optimal performance of quantum conferencing in this type of
protocol is shown in Supplementary Figure~\ref{figS}(a,b). From
Supplementary Figure~\ref{figS}(a), we see that $100$ Bobs within
a radius of $200$m from the relay may extract a key rate at about
0.1 bits per use.

\section{Supplementary Note 5: MDI-QKD star network with discrete-variable
systems}

Let us now discuss a discrete-variable version of the MDI-QKD star
network. In particular, we here show that linear optical
implementations of this network lead to arbitrarily small rates as
the number $N$ of remote users increases, no matter how distant
these users are. In particular, we provide two basic designs which
have two different types of detection at the untrusted relay, with
different costs in terms of bits of CCs. In the entanglement-based
representation, suppose that each of the $N$ Bobs has a Bell pair
$\Phi _{AB}:=\left\vert \Phi\right\rangle _{AB}\left\langle
\Phi\right\vert $ with $\left\vert \Phi\right\rangle :=(\left\vert
00\right\rangle +\left\vert 11\right\rangle )/\sqrt{2}$. Bob keeps
the $B$-qubit while sending the $A$-qubit to the untrusted relay
(since we are interested in an upper bound, we assume there is no
loss and noise in the links, which are therefore identity
channels). Then, the relay detects all the received $A$-qubits in
order to generate an $N$-qubit GHZ state $\left\vert \text{GHZ}_{N}%
^{+}\right\rangle :=(\left\vert 00\ldots0\right\rangle +\left\vert
11\ldots1\right\rangle )/\sqrt{N}$ for the remote $B$-qubits.

\subsection{First design}

In a first design, the relay prepares $N$ local qubits
$C_{1}\ldots C_{N}$ in the state $\left\vert
\text{GHZ}_{N}^{+}\right\rangle $. Then, each $C_{i} $-qubit is
measured with a corresponding $A_{i}$-qubit in a standard qubit
Bell detection, with $4$ possible outcomes. After communicating
$2$ classical bits to the $i$th Bob, the latter can implement a
local Pauli correction, so that the state of the $C_{i}$-qubit is
perfectly teleported to the remote $B_{i}$-qubit. Globally, the
$N$ Bell detections and the $2N$ bits of CC allow one to swap
$\left\vert \text{GHZ}_{N}^{+}\right\rangle $ in the remote
$B$-qubits. The problem appears when we want to implement this
scheme practically with linear optics. In fact, any linear optical
realization of a qubit Bell detection has $1/2$ probability of
success~\cite{Bell1,Bell2,telereview}. The probability of having
$N$
successful Bell detections, and therefore swapping $\left\vert \text{GHZ}%
_{N}\right\rangle $, is therefore $p_{\text{succ}}=2^{-N}$.
Clearly this prevents to share an $N $-partite conference key for
large $N$, no matter at what distance the Bobs are.

\subsection{Second design}

We may consider a different joint measurement at the untrusted
relay. Given
$\left\vert \text{GHZ}_{N}^{+}\right\rangle $ and $\left\vert \text{GHZ}%
_{N}^{-}\right\rangle :=(\left\vert 00\ldots0\right\rangle
+\left\vert 11\ldots1\right\rangle )/\sqrt{N}$, one may construct
a basis of $2^{N}$ orthonormal states by applying the identity $I$
and bit-flip Pauli operators $X$ to $\left\vert
\text{GHZ}_{N}^{\pm}\right\rangle $. For instance, for
$N=3$, we may build the $8$ states (up to normalization)%
\begin{align}
I\otimes I\otimes I\left\vert \text{GHZ}_{3}^{\pm}\right\rangle  &
=\left\vert 000\right\rangle \pm\left\vert 111\right\rangle ,\\
I\otimes I\otimes X\left\vert \text{GHZ}_{3}^{\pm}\right\rangle  &
=\left\vert 001\right\rangle \pm\left\vert 110\right\rangle ,\\
I\otimes X\otimes I\left\vert \text{GHZ}_{3}^{\pm}\right\rangle  &
=\left\vert 010\right\rangle \pm\left\vert 101\right\rangle ,\\
X\otimes I\otimes I\left\vert \text{GHZ}_{3}^{\pm}\right\rangle  &
=\left\vert 100\right\rangle \pm\left\vert 011\right\rangle .
\end{align}
For $N=4$, we may build the $2^{4}=16$ states%
\begin{align}
I\otimes I\otimes I\otimes I\left\vert
\text{GHZ}_{4}^{\pm}\right\rangle  &
=\left\vert 0000\right\rangle \pm\left\vert 1111\right\rangle ,\\
I\otimes I\otimes I\otimes X\left\vert
\text{GHZ}_{4}^{\pm}\right\rangle  &
=\left\vert 0001\right\rangle \pm\left\vert 1110\right\rangle \\
I\otimes I\otimes X\otimes I\left\vert
\text{GHZ}_{4}^{\pm}\right\rangle  &
=\left\vert 0010\right\rangle \pm\left\vert 1101\right\rangle ,\\
I\otimes X\otimes I\otimes I\left\vert
\text{GHZ}_{4}^{\pm}\right\rangle  &
=\left\vert 0100\right\rangle \pm\left\vert 1011\right\rangle ,\\
X\otimes I\otimes I\otimes I\left\vert
\text{GHZ}_{4}^{\pm}\right\rangle  &
=\left\vert 1000\right\rangle \pm\left\vert 0111\right\rangle ,\\
I\otimes I\otimes X\otimes X\left\vert
\text{GHZ}_{4}^{\pm}\right\rangle  &
=\left\vert 0011\right\rangle \pm\left\vert 1100\right\rangle ,\\
I\otimes X\otimes I\otimes X\left\vert
\text{GHZ}_{4}^{\pm}\right\rangle  &
=\left\vert 0101\right\rangle \pm\left\vert 1010\right\rangle ,\\
X\otimes I\otimes I\otimes X\left\vert
\text{GHZ}_{4}^{\pm}\right\rangle  & =\left\vert 1001\right\rangle
\pm\left\vert 0110\right\rangle .
\end{align}
For generic $N\geq3$, the $2^{N}$ basis states are built by
considering all the various combinations in which we can apply the
bit-flip operators. The orthogonal projectors onto this basis\
$\{\Pi_{1},\ldots,\Pi_{k},\ldots ,\Pi_{2^{N}}\}$ provide a von
Neumann measurement. After the relay performs this measurement on
all the $A$-qubits, it broadcasts the $N$\ classical bits of the
outcome $k$ to all Bobs. The latter may therefore apply $X$ or $I
$ operators on their $B$-qubits so that they share the state
$\left\vert \text{GHZ}_{N}^{\pm}\right\rangle $, which can always
be transformed into $\left\vert \text{GHZ}_{N}^{+}\right\rangle $
by means of a single phase-flip Pauli operator $Z$. Again the
problem appears when we want to implement this scheme practically
with linear optics. In fact, as shown in Ref.~\cite{GHZlinear}, a
GHZ-state analyzer based on linear optics can only distinguish two
among $2^{N}$ maximally-entangled GHZ states. This means that the
probability of success is equal to $p_{\text{succ}}=2^{1-N}$. It
is clear that the use of non-linearities may improve this
performance~\cite{GHZnonlinear}, but the scheme rapidly becomes
highly impractical for increasing $N$. Similarly, there is no room
for improving the performance by resorting to
hyper-entanglement~\cite{GHZhyper}, just because the protocol is
strictly based on a single degree of freedom (e.g., polarization).

\section{Supplementary Note 6: Study of imperfections for a proof-of-principle
experiment}

Here we study the impact of practical imperfections on the
performance of our scheme, i.e., on the key rate and achievable
distance within each module $M_{i}$. The degradation of
performances are caused by optical losses that, for instance, may
arise from anti-reflection coating of the beam splitters, and
imperfect beam interferences. Optical losses impact on the
performance mainly because they degrade the operation realized by
the relay. These are analyzed including non-ideal efficiencies of
the relays' detectors (Case 1), and we also consider the combined
effect of detection efficiency and imperfect beam interference
(Case 2).

\subsection{Case 1}

Optical losses are simulated by placing $N$ identical beam
splitters with transmissivity $\tilde{\eta}$ before the homodyne
detections realized by the relay.\begin{figure*}[pth]
\par
\begin{center}
\includegraphics[width=0.95\textwidth]{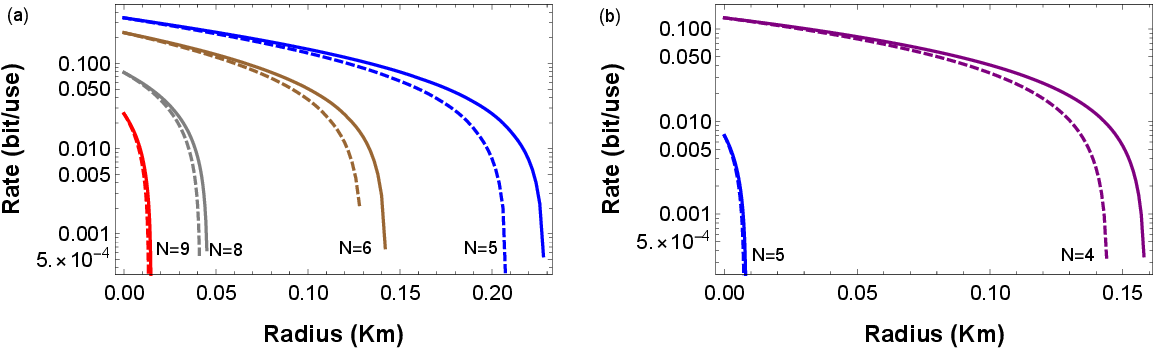}
\end{center}
\par
\vspace{-0.0cm}\caption{This figure shows the optimal key-rate
versus the achievable distance. In panel (a) we plot the key rates
for detector efficiency $\tilde{\eta}=99\%$, considering $N=5$
(blue), $6$ (brown), $8$ (gray), $9$ (red) users, pure-loss
attacks ($\omega=1$, solid), and thermal-loss attacks with
$n_{th}=0.05$ (dashed). The curves in panel (b) are obtained
considering $\tilde{\eta}=98\%$, for $N=4$ (purple) and $5$ (blue).}%
\label{rate-detectors-eff}
\end{figure*} It is important to stress that
the presence of the detector efficiencies $\tilde{\eta}$ preserves
the symmetric structure of the CM in Eq.~(\ref{V2sym}) (see, for
instance, the case with two-users discussed in ref.~\cite{PRA-S}).
In such a case, the only change occurring is in the quantity $x$
of Eq.~(\ref{xyzDEF}), which can be redefined as follows%
\begin{equation}
\tilde{x}:=\eta\mu+(1-\eta)\omega+\frac{1-\tilde{\eta}}{\tilde{\eta}%
}.\label{xbar}%
\end{equation}
Keeping the symmetric structure of the CM (\ref{V2sym}) allows us
to obtain analytical results. In particular, it is simple to see
that the $N$-fold degenerate total symplectic eigenvalue of
Eq.~(\ref{SI-NI-T}) acquires now the following expression
\begin{equation}
\tilde{\nu}=\sqrt{y\left(  y-\frac{z^{2}}{\tilde{x}}\right)  }%
,\label{ni-TOT-tilde}%
\end{equation}
while the conditional spectrum is given by $\tilde{\nu}$,
$(N-2)$-fold
degenerate, and by%
\begin{equation}
\bar{\nu}_{N}=\sqrt{\frac{\left[  N\tilde{\omega}\mu+\eta\left(
1+\left( N-1-N\omega\right)  \mu\right)  \right]  \left[
N\tilde{\omega}\mu +\eta\left[  N-1-\left(  N\omega-1\right)
\mu\right]  \right]  }{\left[
N\tilde{\omega}+\eta(N-1+\mu-N\omega)\right]  \left[  N\tilde{\omega}%
+\eta\left[  1+(N-1)\mu-N\omega\right]  \right]  }},\label{ni-CON-tilde}%
\end{equation}
which is the same as the eigenvalue of Eq.~(\ref{nik}) with
$\tilde{\omega}:=\omega+$ $\left( 1-\tilde{\eta}\right)
/\tilde{\eta}$. The eigenvalues of Eq.~(\ref{ni-TOT-tilde}) and
(\ref{ni-CON-tilde}) are then used to compute the Holevo function
$\tilde{\chi}(\beta_{i},\mathbf{E})$ using the formal expression
of Eq.~(\ref{chi}) and replacing $\nu\rightarrow\tilde{\nu}$ and
$\nu_{N}\rightarrow\bar{\nu}_{N}$. Similarly, the mutual
information between two arbitrary Bobs is obtained from the
two-mode CM $\mathbf{\tilde {V}}_{B_{i}B_{j}}$, having the same
structure of CM $\mathbf{V}_{B_{i}B_{j}}$ of Eq.~(\ref{V2}),
replacing $x\rightarrow\tilde{x}$. Using the following
definitions%
\begin{align}
a &  :=y-\frac{N-1}{N}\frac{z^{2}}{\tilde{x}},\\
b &  :=y-\frac{z^{2}}{N\tilde{x}},\\
c &  :=\frac{z^{2}}{N\tilde{x}},
\end{align}
we write the general expression of the mutual information in the
following
compact form%
\begin{equation}
\tilde{I}(\beta_{i}:\beta_{j})=\frac{1}{2}\log_{2}\frac{\left(
a+1\right) ^{2}\left(  b+1\right)  ^{2}}{\left[  \left( 1+a\right)
^{2}-c^{2}\right]
\left[  \left(  1+b\right)  ^{2}-c^{2}\right]  },\label{I-tilde}%
\end{equation}
and obtain the key rate%
\begin{equation}
\tilde{K}_{ij}(\mu,\eta,\omega,N,\tilde{\eta}):=\tilde{I}(\beta_{i}:\beta
_{j})-\tilde{\chi}(\beta_{i},\mathbf{E}).\label{K-losses-1}%
\end{equation}
The final key-rate, for fixed $\eta,\omega,N,$ and $\tilde{\eta}$,
is obtained
by maximizing over modulation $\mu$%
\begin{equation}
\tilde{K}_{opt}=\max_{\mu}\tilde{K}_{ij}(\mu,\eta,\omega,N,\tilde{\eta}).
\end{equation}
Using previous results, we studied the performance of a star
network for several values of the number of users $N$. The
presence of optical losses degrades the achievable distance, the
key-rate and the number of users composing the star network. The
results are described in Supplementary
Figure~\ref{rate-detectors-eff}, where we plot the optimal key
rate considering pure-loss attacks ($\omega=1$, solid lines), and
symmetric thermal-loss attacks with mean thermal photon number
$\bar{n}=0.05$ (dashed lines). In Supplementary
Figure~\ref{rate-detectors-eff} (a) we show several key rates for
increasing number of users, from right to left. We find that, for
state-state-of-the art detectors~\cite{furusawa-detector} with
efficiency $\tilde{\eta}=0.99$, a star network with $N=9$
components may generate secure keys within a radius of about $14$
m. In Supplementary Figure~\ref{rate-detectors-eff} (b) we see the
impact of slightly worse detectors $(\tilde{\eta}=0.98)$. In such
a case, for $N=5$ users, the achievable distance reduces to $8$ m.

\subsection{Case 2}

We consider now the additional optical losses caused by imperfect
beam interference in the beam splitters $T_{k}$ of the
relay.\begin{figure*}[th]
\par
\begin{center}
\includegraphics[width=0.6\textwidth]{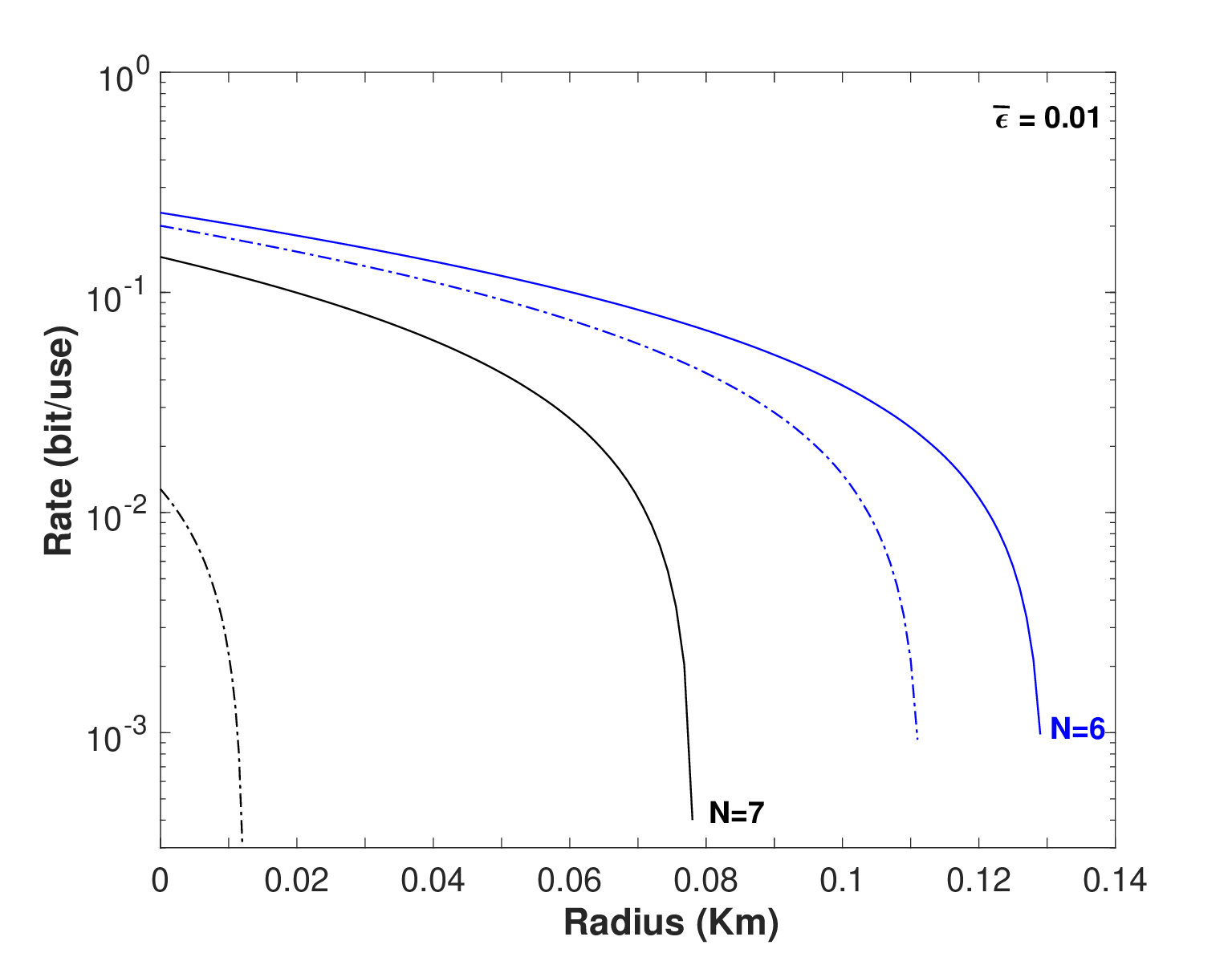}
\end{center}
\par
\vspace{-0.0cm}\caption{This figure shows the optimal key-rate
versus the achievable distance, to assess the combined effect of
non-ideal detector efficiency $\tilde{\eta}=99\%$ and losses
caused by imperfect beam interference (described by setting
$\bar{\epsilon}=1\%$). We plot the optimal key rate of
Eq.~(\ref{Kopt-eps}) for $N=7$ (black), $6$ (blue), assuming
thermal-loss attacks with $n_{th}=0.05$ and $\bar{\epsilon}=0$
(solid lines) or $0.01$ (dot-dashed). We find that the achievable
distance for $N=7$ users reduces to about $12$ m, while for $N=6$
the achievable distance from the
relay is $111$ m.}%
\label{rate-detect-eps}%
\end{figure*} In
addition to the efficiency of the homodyne detectors, we then
consider beam splitters with transmissivity defined as follows
\begin{equation}
T_{k}=1-\frac{1}{k}-\bar{\epsilon}\text{, for }k=2,\dots,N,
\end{equation}
where the parameter $\bar{\epsilon}$ describes the imperfect
beam-splitter interference. We remark that the inclusion of
parameter $\bar{\epsilon}$ breaks the symmetry of the CM that does
not have anymore the structure of Eq.~(\ref{V2sym}). In such a
case, the key rate can be computed only numerically, and
is defined as follows%
\begin{equation}
\bar{K}_{opt}=\inf_{\{i,j\}}\left[
\max_{\mu}\tilde{K}_{ij}(\mu,\eta
,\omega,N,\tilde{\eta},\bar{\epsilon})\right]  ,\label{Kopt-eps}%
\end{equation}
maximizing the key rates $\tilde{K}_{ij}$ between two Bobs over
the Gaussian modulation $\mu$, and taking the infimum over the
pairs $\{i,j\}$. The results of this analysis are described in
Supplementary Figure \ref{rate-detect-eps}. The impact of
parameter $\bar{\epsilon}$ results in a further reduction of the
achievable distance that, for $\bar{\epsilon}=1\%$ and $N=7$,
reduces to $12$ m, while for $N=6$ it is still $111$m. We stress
that, despite we find this degradation of the performance imposed
by practical imperfections, larger number of users can be
connected over arbitrary distances using the modular structure
discussed in this work.

The key rates described in Supplementary
Figures~\ref{rate-detectors-eff} and~\ref{rate-detect-eps} are
obtained from the asymptotic analysis and assuming an imperfect
interferometer. Because our protocol is based on Gaussian
modulation, the optimal attack is Gaussian and, without loss of
generality, the CM formalism can be used to study the security
performance. To determine the entries of the CM, one can use long
data blocks ($n\geq 10^{7}$) for the parameter estimation
procedure, so as to achieve a precise estimation of attenuation
and noise in the channels. In addition to this, the hybrid
(classical/quantum) structure of our modular protocol avoids the
degradation of the key in its propagation throughout the network,
after the completion of quantum communication within each module.

Further analysis may be required to include the impact of
additional imperfections which may play a role in a practical
implementation of the scheme, especially for the case described in
Supplementary Note 4. In such a case it may be necessary to
consider the multimode structure of the sources and the local
detectors, which may destabilize the achievable key rate and the
security properties, as described in Ref.~\cite{usenko-multimode}
for one-way protocols. However, the security properties of the
single-mode analysis can be recovered by applying a suitable mode
symmetrization that neutralizes the leakage of information towards
the eavesdropper~\cite{usenko-multimode}.

\end{document}